\documentclass[aps,prb,letterpaper,showpacs]{revtex4}
\usepackage{keyval}%
\usepackage{graphicx}
\usepackage{dcolumn}
\usepackage{bm}
\usepackage{color}
\usepackage[intlimits]{amsmath}

\setkeys{Gin}{width=0.8\columnwidth,clip=true}

\newcommand{\ignore}[1]{}

\begin{document}


\title{Flavor-twisted boundary condition for simulations of quantum many-body systems}


\author{Wei-Guo Yin} 
\email[Corresponding author. To whom all correspondence should be
addressed. E-mail: ]{wyin@bnl.gov}
\author{Wei Ku}
\affiliation{Condensed Matter Physics and Materials Science Department, Brookhaven National Laboratory, Upton, New York 11973, U.S.A.} %

\date{Received \today}

\begin{abstract}
We present an approximative simulation method for quantum many-body
systems based on coarse graining the space of the momentum
transferred between interacting particles, which leads to effective
Hamiltonians of reduced size with the flavor-twisted boundary
condition. A rapid, accurate, and fast convergent computation of the
ground-state energy is demonstrated on the spin-$\frac{1}{2}$
quantum antiferromagnet of any dimension by employing only two
sites. The method is expected to be useful for future simulations
and quick estimates on other strongly correlated systems.
\end{abstract}

\maketitle


Understanding a quantum many-body system is fundamentally
challenging because of the exponential growth of the number of
states with the system size. To estimate the physical properties of
an intractably large system, a common practice is to extrapolate the
results for several significantly reduced
sizes,\cite{HTC:review:dagotto_94} together with the periodic
boundary condition (PBC).\cite{PBC:ashcroft}  A different yet
complementary approach is to continuously twist the boundary
conditions for one solvable size. The typical flavor-independent
version of the latter has been seen in solid state
physics,\cite{TBC:poilblanc,TBC:tohyama} while the flavor version
has been used in quantum
chromodynamics.\cite{FTBC:QCD:deDivitiis,FTBC:QCD:sachrajda,FTBC:QCD:tiburzi}
Here we apply the flavor-twisted boundary conditions
(FTBC)\cite{FTBC:QCD:deDivitiis,FTBC:QCD:sachrajda,FTBC:QCD:tiburzi}
to the spin-$\frac{1}{2}$ quantum antiferromagnet, one of the basic
models in solid state physics. We show that the ground-state energy
can be accurately calculated using only \emph{two} sites.

We begin with an explicit derivation of FTBC, which is necessary for
systematical studies of this and related methods. Let us consider
the connection between a large system of size $L$ and a small one of
size $l$, both with PBC. (To distinguish them, the upper-case and
lower-case letters are used thoroughly for the large and small
lattices, respectively.) The large system is described by the
following general Hamiltonian in the second quantized language and
in the real space representation,
\begin{equation}
H_{L} = \sum_{\mathbf{I},\mathbf{J},\sigma} {(t_{\mathbf{I}\mathbf{J}}^{\sigma} C_{\mathbf{I}\sigma}^\dag
C_{\mathbf{J}\sigma}^{} + h.c.})
+ \sum_{\substack{\mathbf{I},\mathbf{J},\sigma\\ \mathbf{J}^\prime,\mathbf{I}^\prime,\sigma^\prime}} {%
U^{\sigma,\sigma^\prime}_{\mathbf{I}\mathbf{J}^\prime\mathbf{J}\mathbf{I}^\prime}
C_{\mathbf{I}\sigma}^\dag C_{\mathbf{J}^\prime\sigma^\prime}^\dag
C_{\mathbf{I}^\prime\sigma^\prime}^{} C_{\mathbf{J}\sigma}^{}},
\label{eq:model_i}
\end{equation}
where $C_{\mathbf{I}\sigma}^{}$ denote the quantum operator that
annihilates a particle with flavor $\sigma$ at site $\mathbf{I}$.
$L=\prod_{\alpha=1}^{d} {L_\alpha}$ is the total number of the
lattice sites in a $d$-dimensional space with $L_\alpha$ being the
site number in the $\alpha$-dimension. The thermodynamic limit is
reached when all $L_\alpha \to \infty$.

There also exists a reciprocal space where the counterpart of the
site $\mathbf{I}$ is the momentum point $\mathbf{K}$. Imposing PBC
to the real space discretizes the $\mathbf{K}$ space as follows
\cite{PBC:ashcroft}
\begin{equation}
K_{\alpha} = \frac {2\pi}{L_\alpha} M, \;\;\; M=0, 1, 2, \dots,
L_\alpha-1 \label{eq:PBC}
\end{equation}
where all the lattice constants have been scaled to unit.
Eq.~(\ref{eq:PBC}) translates the concepts of \emph{large and small}
in the real space to those of \emph{fine and coarse} in the
$\mathbf{K}$ grid, respectively. In the
reciprocal space, %
\begin{equation}
H_{L}=\sum_{\mathbf{K} \sigma} {\varepsilon_{\mathbf{K}}^{\sigma}
C_{\mathbf{K}\sigma}^\dag C_{\mathbf{K}\sigma}^{}}
+ \frac{1}{L}\sum_{\mathbf{Q}}\{\sum_{\mathbf{K}\sigma,\mathbf{K}^\prime\sigma^\prime}
{U^{\sigma,\sigma^\prime}_{\mathbf{K},\mathbf{K}^\prime,\mathbf{Q}}
C_{\mathbf{K}+\mathbf{Q},\sigma}^\dag
C_{\mathbf{K}^\prime-\mathbf{Q},\sigma^\prime}^\dag
C_{\mathbf{K}^\prime,\sigma^\prime}^{} C_{\mathbf{K},\sigma}^{}}\}
\label{eq:model}
\end{equation}
where $C_{\mathbf{K}\sigma}^{}$ annihilates a particle with momentum
$\mathbf{K}$ and spin $\sigma$ and is given by
\begin{equation}
C_{\mathbf{K}\sigma}^{}=\frac{1}{\sqrt{L}}\sum_{\mathbf{I}}
{e^{-i\mathbf{K}\cdot\mathbf{R}_\mathbf{I}}C_{\mathbf{I}\sigma}^{}},
\label{eq:c}
\end{equation}
where $\mathbf{R}_\mathbf{I}$ is the coordinates of the
$\mathbf{I}$-th site. The bare energy dispersion and the interaction
function are
\begin{eqnarray}
\varepsilon_{\mathbf{K}}^{\sigma} &=&
\sum_{\mathbf{R}_\mathbf{J}-\mathbf{R}_\mathbf{I}}
t_{\mathbf{I}\mathbf{J}}^{\sigma}e^{i\mathbf{K}\cdot(\mathbf{R}_\mathbf{J}-\mathbf{R}_\mathbf{I})}. \\
U^{\sigma,\sigma^\prime}_{\mathbf{K},\mathbf{K}^\prime,\mathbf{Q}} &=&
\sum_{\substack{\mathbf{R}_\mathbf{J}-\mathbf{R}_\mathbf{I}\\
\mathbf{R}_{\mathbf{I}^\prime}-\mathbf{R}_{\mathbf{J}^\prime}\\
\mathbf{R}_{\mathbf{J}^\prime}-\mathbf{R}_\mathbf{I}}}
U^{\sigma,\sigma^\prime}_{\mathbf{I}\mathbf{J}^\prime\mathbf{J}\mathbf{I}^\prime}
e^{i\mathbf{K}\cdot(\mathbf{R}_\mathbf{J}-\mathbf{R}_\mathbf{I})}
e^{i\mathbf{K}^\prime\cdot(\mathbf{R}_{\mathbf{I}^\prime}-\mathbf{R}_{\mathbf{J}^\prime})}
e^{i\mathbf{Q}\cdot(\mathbf{R}_{\mathbf{J}^\prime}-\mathbf{R}_\mathbf{I})}.
\label{eq:U}
\end{eqnarray}

\begin{figure}[t]
\begin{center}
\includegraphics[width=0.6\columnwidth,clip=true]{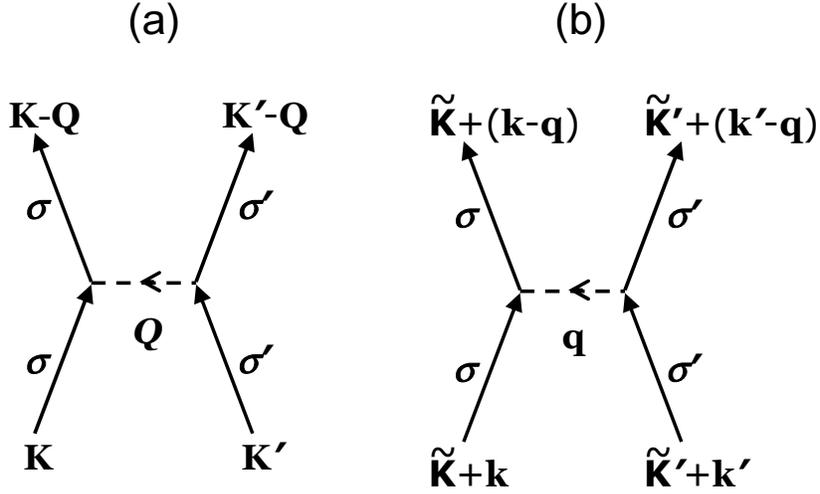}%
\caption{\label{fig:q}%
Illustration of the momentum transfer due to the particle-particle
interaction: (a) original and (b) coarsened.}
\end{center}
\end{figure}

Eq.~(\ref{eq:model}) appears to describe two particles with
$\mathbf{K}$ and $\mathbf{K}^\prime$, respectively, interact with
internal momentum transfer $\mathbf{Q}$, as diagrammed in
Fig.~\ref{fig:q}(a). Since the $\mathbf{Q}$ points are to be
integrated [c.f. $\sum_\mathbf{Q}\{\cdots\}$ in Eq.
(\ref{eq:model})], here comes a well-known numerical trick:
\emph{the summation (integration) over a fine grid may be well
approximated by that over a rather coarse grid.} This numerical
recipe receives particular support in the quantum many-body systems
of interest, where a small number of $\mathbf{Q}$ are far more
important than the others, the so-called $\mathbf{Q}$-mode
resonance, e.g., $\mathbf{Q}=(\pi,\pi,\cdots,\pi)$ in
antiferromagnetic correlation. The $\mathbf{Q}$-mode resonance
implies dramatic response to small stimuli such as changing
temperature, applied fields, pressure, doping, etc, which is
important to determining the functionalities of the system.
Therefore, it may be a good approximation to coarsen the
$\mathbf{Q}$ grid as long as the important $\mathbf{Q}$ points are
included.

To find a way to coarsen the $\mathbf{Q}$ grid such that the
resulting Hamiltonian can be readily transformed back to the real
space (where the parameters are often short-ranged, meriting for
numerical computation), we consider the small system of size $l$
with PBC whose momenta $\mathbf{k}$ are given by:
\begin{equation}
k_{\alpha} = \frac {2\pi}{l_\alpha} m, \;\;\; m=0, 1, 2, \dots,
l_\alpha-1. \label{q}
\end{equation}
where $l=\Pi_{\alpha=1}^{d} {l_\alpha} \ll L$ is the total site
number of the smaller lattice. Let the small system be commensurate
with the large one, that is, all the $\mathbf{k}$ points can be
found in the $\mathbf{K}$ grid of the large system. Then, any
momentum in the fine $\mathbf{K}$ grid can be rewritten as,
\begin{equation}
\mathbf{K}=\tilde{\mathbf{K}}+\mathbf{k}, \;\;\;\;
\mathbf{K}^\prime=\tilde{\mathbf{K}}^\prime+\mathbf{k}^\prime, \;\;\;\;
\mathbf{Q}=\tilde{\mathbf{Q}}+\mathbf{q},
\end{equation}
where $\mathbf{k}^\prime$ and $\mathbf{q}$ are in the $\mathbf{k}$
grid defined in Eq.~(\ref{q}); the $L/l$ `twists'
$\tilde{\mathbf{K}}$ (likewise $\tilde{\mathbf{K}}^\prime$ or
$\tilde{\mathbf{Q}}$) consist of a subset of the original
$\mathbf{K}$ points that fill the interstitial region of the
$\mathbf{k}$ points nearby the origin. For example, for $d=1$,
\begin{equation}
\tilde{K}_{\alpha} \in (-\frac {\pi}{l_\alpha}, \frac
{\pi}{l_\alpha}].
\end{equation}
As another example, $l=2$ for the square lattice means that
$\mathbf{k}=(0,0)$ and $(\pi,\pi)$, and that $\tilde{\mathbf{K}}$
cover 
half of the original $\mathbf{K}$ space satisfying
$\cos(\tilde{K}_x)+\cos(\tilde{K}_y)\geq 0$.

One can explicitly rewrite $H_L$ in terms of the real-space $l$-site
lattice by introducing a new set of particle annihilation operators,
\begin{eqnarray}
d_{\mathbf{k}\sigma;\tilde{\mathbf{K}}}^{}=C_{\tilde{\mathbf{K}}+\mathbf{k}\sigma}^{}=C_{\mathbf{K}\sigma}^{},
\label{eq:d} \\
d_{\mathbf{i}\sigma;\tilde{\mathbf{K}}}^{}=\frac{1}{\sqrt{l}}\sum_{\mathbf{k}}
{e^{i\mathbf{k}\cdot\mathbf{r}_i}d_{\mathbf{k}\sigma;\tilde{\mathbf{K}}}^{}},
\label{eq:d_i}
\end{eqnarray}
where $\mathbf{r}_i$ is the coordinates of the $i$-th site of this
$l$-site lattice, satisfying
\begin{equation}
\sum_{\mathbf{k}}
e^{i\mathbf{k}\cdot(\mathbf{r}_\mathbf{j}-\mathbf{r}_\mathbf{i})} =
l\delta_{\mathbf{i},\mathbf{j}}. \label{eq:sum}
\end{equation}
By using Eqs.~(\ref{q}-\ref{eq:sum}), the original Hamiltonian for
the $L$-site system, Eq. (\ref{eq:model_i}) or Eq.~(\ref{eq:model}),
can be transformed to a set of $l$-site subsystems described by the
following new Hamiltonian in the real space representation
\begin{eqnarray}
H_L &=& \sum_{\tilde{\mathbf{K}}\sigma}
  \sum_{\mathbf{i},\mathbf{j}}
    [e^{i\tilde{\mathbf{K}}\cdot(\mathbf{r}_j-\mathbf{r}_i)}t_\mathbf{ij}^{\sigma}
d_{\mathbf{i}\sigma;\tilde{\mathbf{K}}}^\dag
d_{\mathbf{j}\sigma;\tilde{\mathbf{K}}}^{}+h.c.] + %
\frac{l}{L}\sum_{\tilde{\mathbf{Q}}}\;
\sum_{\tilde{\mathbf{K}}\sigma,\tilde{\mathbf{K}}^\prime\sigma^\prime}\;
\sum_{\mathbf{i},\mathbf{i}^\prime,\mathbf{j},\mathbf{j}^\prime}
\nonumber \\
&& { \left\{
e^{i\tilde{\mathbf{K}} \cdot (\mathbf{r}_j-\mathbf{r}_i)} %
e^{i\tilde{\mathbf{K}}^\prime \cdot
(\mathbf{r}_{i^\prime}-\mathbf{r}_{j^\prime})}
e^{i\tilde{\mathbf{Q}} \cdot (\mathbf{r}_{j^\prime}-\mathbf{r}_{i})}
U_{\mathbf{i}\mathbf{j}^\prime\mathbf{j}\mathbf{i}^\prime}^{\sigma,\sigma^\prime}
d_{\mathbf{i}\sigma;\tilde{\mathbf{K}}+\tilde{\mathbf{Q}}}^\dag
d_{\mathbf{j}^\prime\sigma^\prime;\tilde{\mathbf{K}}^\prime-\tilde{\mathbf{Q}}}^\dag
d_{\mathbf{i}^\prime\sigma^\prime;\tilde{\mathbf{K}}^\prime}^{}
d_{\mathbf{j}\sigma;\tilde{\mathbf{K}}}^{} \right\} }.
\label{eq:model_d_0}
\end{eqnarray}
This establishes the exact transformation between the large and the
small, subject to the linear size of the small system being not
shorter than the range of any Hamiltonian parameters of Eq.
(\ref{eq:model_i}).

\emph{Approximation 1.---}Now we coarse the $\mathbf{Q}$ grid with
\begin{equation}
\tilde{\mathbf{Q}}\equiv 0. \label{eq:Q=0}
\end{equation}
This means that only the $\mathbf{q}$ points, whose total number is
$l$, are retained as the internal momentum transfer. Then,
$\mathbf{K} + \mathbf{q} = \tilde{\mathbf{K}}+\mathbf{k} +
\mathbf{q} = \tilde{\mathbf{K}}+\mathbf{k}^\prime $. That is, any
$\mathbf{K}$ point is scattered to the sum of the same twist and
another point in $\{\mathbf{q}\}$. Thus, \emph{a particle with a
given $\tilde{\mathbf{K}}$ can be scattered by the particle-particle
interaction into only $l$ points in the $\mathbf{K}$ momentum space,
instead of $L$ points}, as illustrated in Fig.~\ref{fig:q}(b).
Eq.~(\ref{eq:model_d_0}) is rewritten as
\begin{equation}
H_L \simeq
\frac{l}{L}\sum_{\tilde{\mathbf{K}}\sigma,\tilde{\mathbf{K}}^\prime\sigma^\prime}
 H_{l}({\tilde{\mathbf{K}}\sigma,\tilde{\mathbf{K}}^\prime\sigma^\prime}),
 \label{eq:l2s}
\end{equation}
where the subsystems are
\begin{eqnarray}
  H_{l}({\tilde{\mathbf{K}}\sigma,\tilde{\mathbf{K}}^\prime\sigma^\prime})
  &=& \sum_{\mathbf{i},\mathbf{j}}
    [e^{i\tilde{\mathbf{K}}\cdot(\mathbf{r}_j-\mathbf{r}_i)}t_\mathbf{ij}^{\sigma}
d_{\mathbf{i}\sigma;\tilde{\mathbf{K}}}^\dag
d_{\mathbf{j}\sigma;\tilde{\mathbf{K}}}^{}+h.c.] + \nonumber \\
  && + \sum_{\mathbf{i},\mathbf{j}}
    [e^{i\tilde{\mathbf{K}}^\prime\cdot(\mathbf{r}_j-\mathbf{r}_i)}t_\mathbf{ij}^{\sigma^\prime}
d_{\mathbf{i}\sigma^\prime;\tilde{\mathbf{K}}^\prime}^\dag
d_{\mathbf{j}\sigma^\prime;\tilde{\mathbf{K}}^\prime}^{}+h.c.] + \nonumber \\
&& + \sum_{\mathbf{i},\mathbf{j},\mathbf{j}^\prime,\mathbf{i}^\prime} {%
e^{i\tilde{\mathbf{K}} \cdot (\mathbf{r}_j-\mathbf{r}_i)} %
e^{i\tilde{\mathbf{K}}^\prime \cdot
(\mathbf{r}_{i^\prime}-\mathbf{r}_{j^\prime})}
U_{\mathbf{i}\mathbf{j}^\prime\mathbf{j}\mathbf{i}^\prime}^{\sigma,\sigma^\prime}
d_{\mathbf{i}\sigma;\tilde{\mathbf{K}}}^\dag
d_{\mathbf{j}^\prime\sigma^\prime;\tilde{\mathbf{K}}^\prime}^\dag
d_{\mathbf{i}^\prime\sigma^\prime;\tilde{\mathbf{K}}^\prime}^{}
d_{\mathbf{j}\sigma;\tilde{\mathbf{K}}}^{}}. \label{eq:model_d_i}
\end{eqnarray}
Combined with Eq.~(\ref{eq:l2s}), Eq.~(\ref{eq:model_d_i}) for each
pair of $\tilde{\mathbf{K}}$ and $\tilde{\mathbf{K}}^\prime$ is
\emph{not} independent of another pair. For example, the
\{$\tilde{\mathbf{K}}$, $\tilde{\mathbf{K}}^\prime$\} subsystem and
the \{$\tilde{\mathbf{K}}$, $\tilde{\mathbf{K}}^{\prime\prime}$\}
subsystem share the same momentum points with $\tilde{\mathbf{K}}$.
Therefore, all the $(L/l)^2$ subsystems of size $l$ are connected
and one still has to deal with a case of size $L$. To simplify the
calculations to size $l$, a further approximation is needed.

\emph{Approximation 2 (FTBC).---}A simple approximation is to treat
$H_{l}({\tilde{\mathbf{K}}\sigma,\tilde{\mathbf{K}}^\prime\sigma^\prime})$
independently. Thus, to estimate the expectation value of an
observable $\hat{O}_L$ in the large system is to calculate
\begin{equation}
\mathrm{trace}(\hat{\rho}_L \, \hat{O}_L)\simeq
\sum_{\tilde{\mathbf{K}}\sigma,\tilde{\mathbf{K}}^\prime\sigma^\prime} \mathrm{trace}(
\hat{\rho}_{l; {\tilde{\mathbf{K}}\sigma,\tilde{\mathbf{K}}^\prime\sigma^\prime}}\,
\hat{O}_{l; {\tilde{\mathbf{K}}\sigma,\tilde{\mathbf{K}}^\prime\sigma^\prime}}),
\end{equation}
where $\hat{\rho}_L$ denotes the density matrix for the large system
and $\hat{\rho}_{l;
{\tilde{\mathbf{K}}\sigma,\tilde{\mathbf{K}}^\prime\sigma^\prime}}$
denotes the density matrix for the isolated small system with the
twists $ \tilde{\mathbf{K}}$ and $\tilde{\mathbf{K}}^\prime$, as
determined by Eq.~(\ref{eq:model_d_i}). $\hat{O}_{l;
{\tilde{\mathbf{K}}\sigma,\tilde{\mathbf{K}}^\prime\sigma^\prime}}$
is the transformed observable following
Eqs.~(\ref{eq:d})-(\ref{eq:sum}) and Eq.~(\ref{eq:Q=0}). For
example, the ground state energy is approximated by
\begin{equation}
\langle \Phi_L | \sum_{\tilde{\mathbf{K}}\sigma,\tilde{\mathbf{K}}^\prime\sigma^\prime}
 H_{l}({\tilde{\mathbf{K}}\sigma,\tilde{\mathbf{K}}^\prime\sigma^\prime})| \Phi_L \rangle
 \simeq \sum_{\tilde{\mathbf{K}}\sigma,\tilde{\mathbf{K}}^\prime\sigma^\prime}
\langle \phi_{l;{\tilde{\mathbf{K}}\sigma,\tilde{\mathbf{K}}^\prime\sigma^\prime}}
 | H_{l}({\tilde{\mathbf{K}}\sigma,\tilde{\mathbf{K}}^\prime\sigma^\prime}) |
 \phi_{l; {\tilde{\mathbf{K}}\sigma,\tilde{\mathbf{K}}^\prime\sigma^\prime}}
\rangle,
 \label{eq:ftbc}
\end{equation}
where $|\Phi_L \rangle$ denotes the ground-state wave function for
the large system and $|\phi_{l;
{\tilde{\mathbf{K}}\sigma,\tilde{\mathbf{K}}^\prime\sigma^\prime}}
\rangle$ denotes a wave function for the isolated small system with
the twists $ \tilde{\mathbf{K}}$ and $\tilde{\mathbf{K}}^\prime$.

This approximation is actually equivalent to FTBC
(Ref.~\onlinecite{FTBC:QCD:tiburzi}) as explained below. Comparing
Eq.~(\ref{eq:model_d_i}) with Eq.~(\ref{eq:model_i}), one finds that
they are very similar in form---only differ in the parameters by a
phase factor $e^{i\tilde{\mathbf{K}} \cdot
(\mathbf{r}_i-\mathbf{r}_j)}$ associated with the momentum twist
$\tilde{\mathbf{K}}$. Actually, solving Eq.~(\ref{eq:model_d_i}) is
equivalent to solving the following Hamiltonian for the $l$-site
lattice
\begin{eqnarray}
H_{l}({\tilde{\mathbf{K}}\sigma,\tilde{\mathbf{K}}^\prime\sigma^\prime})
 &=& \sum_{\mathbf{i},\mathbf{j}} {(t_\mathbf{ij}^{\sigma} d_{\mathbf{i}\sigma;\tilde{\mathbf{K}}}^\dag
d_{\mathbf{j}\sigma;\tilde{\mathbf{K}}}^{} + h.c.) } %
+ \sum_{\mathbf{i},\mathbf{j}} {(t_\mathbf{ij}^{\sigma^\prime} d_{\mathbf{i}\sigma^\prime;\tilde{\mathbf{K}}^\prime}^\dag
d_{\mathbf{j}\sigma^\prime;\tilde{\mathbf{K}}^\prime}^{} + h.c.) } \nonumber \\
&&+ \sum_{\mathbf{i},\mathbf{j},\mathbf{j}^\prime,\mathbf{i}^\prime} {%
U_{\mathbf{i}\mathbf{i}^\prime\mathbf{j}\mathbf{j}^\prime}^{\sigma,\sigma^\prime}
d_{\mathbf{i}\sigma;\tilde{\mathbf{K}}}^\dag
d_{\mathbf{i}^\prime\sigma^\prime;\tilde{\mathbf{K}}^\prime}^\dag
d_{\mathbf{j}^\prime\sigma^\prime;\tilde{\mathbf{K}}^\prime}^{}
d_{\mathbf{j}\sigma;\tilde{\mathbf{K}}}^{}} \label{eq:model_f}
\end{eqnarray}
together with the boundary conditions such that translating its
many-body wavefunction $| \Psi_l \rangle$ along the $\alpha$-th
dimension $l_\alpha$ steps yield $ e^{i (N_1\tilde{K}_\alpha +
N_2\tilde{K}^\prime_{\alpha}) l_\alpha} | \Psi_l \rangle$ where
$N_1$ and $N_2$ are the numbers of the two flavors of particles
associated with the two twists $\tilde{\mathbf{K}}$ and
$\tilde{\mathbf{K}}^\prime$, respectively.

With FTBC, the problem of solving the original Hamiltonian for a
$L$-site system is reduced to solving $(L/l)^2$ Hamiltonians for
$l$-site subsystems, each corresponding to a given \emph{pair} of
$\tilde{\mathbf{K}}$ and $\tilde{\mathbf{K}}^\prime$. The
computational benefit is substantial, since the number of states
grows exponentially with the system size. For the Hubbard model as
an example, the calculation load is reduced from $O(4^L4^L)$ to
$O[(L/l)^2 \times 4^l4^l]$. In addition, the $(L/l)^2$ prefactor is
fully parallelizable and it can be readily handled by using the same
integration trick of replacing a fine-grid sum with a coarse-grid
sum. Since FTBC is reached at the level of Hamiltonian and
interpreted as boundary conditions, it has least limitation and
fully compatible with other many-body approaches, e.g., Lanczos
exact diagonalization and quantum Monte Carlo
methods.\cite{HTC:review:dagotto_94}

\emph{Approximation 3.---}The typical twisted boundary
conditions\cite{TBC:poilblanc,TBC:tohyama} (TBC) can be reached from
Eqs.~(\ref{eq:l2s}) and (\ref{eq:model_d_i}) by further taking the
$\tilde{\mathbf{K}}=\tilde{\mathbf{K}}^\prime$ approximation,
\begin{equation}
\frac{l}{L}\sum_{{\tilde{\mathbf{K}}\sigma,\tilde{\mathbf{K}}^\prime\sigma^\prime}}
 H_{l}({\tilde{\mathbf{K}}\sigma,\tilde{\mathbf{K}}^\prime}\sigma^\prime)
\approx
\sum_{{\tilde{\mathbf{K}}\sigma,\sigma^\prime}}
H_{l}(\tilde{\mathbf{K}}\sigma,\tilde{\mathbf{K}}\sigma^\prime).
\end{equation}
With TBC, each subsystem has only one momentum twist
$\tilde{\mathbf{K}}$. Commonly, the one-site or two-site
($\mathbf{r}_{i^\prime}=\mathbf{r}_i$ and
$\mathbf{r}_{j^\prime}=\mathbf{r}_j$) interactions are the leading
interaction terms. Then, the twisted phases of the interaction terms
in Eq.~(\ref{eq:model_d_i}) cancel, which renders the interaction
terms for $l$ to be of the same form as those for $L$. Thus, the
approximation by TBC allows the continuous sampling of the momentum
space for one-particle excitations, but it prohibits the same for
two-particle excitations. To compare, FTBC allows both in principle.
It could be expected that FTBC is more accurate, as illustrated
below.

To complete, PBC is an additional approximation of TBC
($\tilde{\mathbf{K}}=\tilde{\mathbf{K}}^\prime=0$). FTBC with one
twist zero fixed ($\tilde{\mathbf{K}}^\prime=0$, referred to as
FTBC0)\cite{FTBC:QCD:deDivitiis,FTBC:QCD:sachrajda} was studied
before.

To illustrate all the above points, we use FTBC to estimate the
ground state energy of the spin-1/2 Heisenberg quantum
antiferromagnet of any dimension. Here the flavor is the spin of
electrons, consisting of $\uparrow$ and
$\downarrow$.\cite{spin:note} The results are compared with those
obtained from using the other boundary conditions and linear
spin-wave theory (LSW).\cite{spin:sw:anderson}

The spin-1/2 antiferromagnetic Heisenberg model is given by
\begin{equation}
H_L=J\sum_{\langle \mathbf{I},\mathbf{J} \rangle}
{[S_\mathbf{I}^{z}S_\mathbf{J}^{z}+\frac{1}{2}(S_\mathbf{I}^{+}S_\mathbf{J}^{-}+S_\mathbf{I}^{-}S_\mathbf{J}^{+})]},
\label{eq:Heisenberg}
\end{equation}
where the spin operator $S_{\mathbf{I}}^{+,-,z}=\sum_{\mu\nu}
{C_{\mathbf{I}\mu}^\dag \sigma^{+,-,z}_{\mu\nu}
C_{\mathbf{I}\nu}^{}}$ with $\sigma^{+,-,z}_{\mu\nu}$ being the
Pauli matrix elements. $\sum_{\langle \mathbf{I},\mathbf{J}
\rangle}$ runs over  nearest neighbors. It is not only the basic
account of antiferromagnetism \cite{spin:mattis} but also the ground
zero of understanding high-temperature superconductivity in copper
oxides.\cite{spin:review:manousakis,HTC:model:anderson} Two
interesting states have been intensively considered in the
literature: the long-range ordered N\'{e}el state and the
resonating-valence-bond (RVB)
state.\cite{HTC:model:anderson,spin:triangular:anderson} The concept
of RVB is based on the fact that the minimum bond energy ($-0.75 J$)
is realized in a two-site system, much lower than the bond energy
($-0.25 J$) of the classic N\'{e}el state; the valence-bond between
two nearest-neighbor spins is arguably key to understanding
low-dimensional quantum
antiferromagnets.\cite{spin:triangular:anderson}  However, the
2-site dimer breaks bonding to spins on the other sites in an
extended system. For example, the total energy from the dimers is
$-0.375 J$ per bond for $d=1$ and $-0.1875 J$ per bond for $d=2$,
much higher than the exact result \cite{spin:bethe,spin:hulthen}
$-0.443 J$ for $d=1$ and the numerical result
\cite{spin:square:huse} $-0.334 J$ for $d=2$, respectively. It is
argued that the resonating (i.e., superposition of the degenerate
states of different dimer configurations) could lower the energy
substantially.\cite{spin:triangular:anderson} But an accurate
estimation of the ground state energy was achieved only when
long-distance spin dimers were also
included.\cite{spin:square:liang}

With FTBC, after coarsening the $\mathbf{Q}$ mesh, one obtains
\begin{equation}
H_{l}({\tilde{\mathbf{K}}_\uparrow,\tilde{\mathbf{K}}_\downarrow}) =
J\sum_{\langle\mathbf{i},\mathbf{j}\rangle} \{ {
s_\mathbf{i}^{z}s_\mathbf{j}^{z}} + \frac{1}{2}[e^{i(\tilde{\mathbf{K}}_\uparrow-\tilde{\mathbf{K}}_\downarrow)\cdot(\mathbf{r}_j-\mathbf{r}_i)}
s_\mathbf{i}^{+}s_\mathbf{j}^{-}+h.c.] \}, \label{eq:Heisenberg_l}
\end{equation}
where
$s_{\mathbf{i}}^{+}=d_{\mathbf{i}\uparrow;\tilde{\mathbf{K}}\uparrow}^\dag
d_{\mathbf{i}\downarrow;\tilde{\mathbf{K}}\downarrow}^{}$,
$s_{\mathbf{i}}^{-}=d_{\mathbf{i}\downarrow;\tilde{\mathbf{K}}\downarrow}^{\dag}d_{\mathbf{i}\uparrow;\tilde{\mathbf{K}}\uparrow}^{}$
and
$s_{\mathbf{i}}^{z}=\frac{1}{2}(d_{\mathbf{i}\uparrow;\tilde{\mathbf{K}}\uparrow}^\dag
d_{\mathbf{i}\uparrow;\tilde{\mathbf{K}}\uparrow}^{}-d_{\mathbf{i}\downarrow;\tilde{\mathbf{K}}\downarrow}^\dag
d_{\mathbf{i}\downarrow;\tilde{\mathbf{K}}\downarrow}^{})$. The
spin-exchange terms strongly depend on the double twists,
$\tilde{\mathbf{K}}_\uparrow$ and $\tilde{\mathbf{K}}_\downarrow$,
for the spin-up and spin-down electrons, respectively. In
comparison, the result with TBC is
\begin{equation}
H_{l}(\tilde{\mathbf{K}}_\uparrow,\tilde{\mathbf{K}}_\downarrow) = J\sum_{\mathbf{i},\mathbf{j}} \{ { s_\mathbf{i}^{z}s_\mathbf{j}^{z}} + \frac{1}{2}(
s_\mathbf{i}^{+}s_\mathbf{j}^{-}+s_\mathbf{i}^{-}s_\mathbf{j}^{+})
\}, \label{eq:Heisenberg_TBC}
\end{equation}
independent of twists, the same as with PBC in this case.

The accuracy and convergency of FTBC are tested with the bond energy
\begin{equation}
E({L;l}) =
\frac{1}{n_\mathrm{bond}}\left(\frac{l}{L}\right)^2\sum_{\tilde{\mathbf{K}}_\uparrow,\tilde{\mathbf{K}}_\downarrow}
 \langle
 H_{l}(\tilde{\mathbf{K}}_\uparrow,\tilde{\mathbf{K}}_\downarrow)\rangle,
\end{equation}
where $n_\mathrm{bond}$ is the number of the nearest-neighbor AF
bonds in the small system. Let us first take the most dramatic
approximation, viz: $l=2$ (note $n_\mathrm{bond}=1$), the
fundamental of the RVB state. With PBC, the $\mathbf{q}$ mesh
contains only two points: $(0,0,\dots,0)$ and $(\pi,\pi,\dots,\pi)$
corresponding to a spin singlet and a triplet, respectively; the
energy of the $\mathbf{q}=0$ state is $-0.75 J$. By using FTBC to
continuously and smoothly twist the energies of the subsystems, the
bond energy becomes
\begin{eqnarray}
E(L;l=2) &=&
J\left(\frac{2}{L}\right)^2\sum_{\tilde{\mathbf{K}}_\uparrow,\tilde{\mathbf{K}}_\downarrow}
{ [-\frac{1}{4} - \frac{1}{2}
e^{i(\tilde{\mathbf{K}}_\uparrow-\tilde{\mathbf{K}}_\downarrow)\cdot(\mathbf{r}_1-\mathbf{r}_0)}]} \nonumber \\
&=& -\frac{J}{4} - \frac{J}{2} \prod_{\alpha=1}^d
\frac{1}{(2\pi)^2}\int_{-\pi}^\pi
\mathrm{d}\tilde{K}^\alpha_\uparrow \int_{-\pi}^\pi
\mathrm{d}\tilde{K}^\alpha_\downarrow
{\cos(\tilde{K}^\alpha_\uparrow /2)\cos(\tilde{K}^\alpha_\downarrow
/2)} \nonumber \\
&=& -\frac{J}{4} - \frac{J}{2} \left(\frac{2}{\pi}\right)^{2d} \;\;
\mathrm{for}\; L \to \infty. \label{eq:e}
\end{eqnarray}
The results are listed in Table~\ref{table:energy}, together with
those obtained from using FTBC0, TBC, and PBC as well as the LSW
results. The bond energy in the thermodynamic limit and for any
dimension is accurately reproduced using the simple Eq.~(\ref{eq:e})
for $l=2$.

\begin{table}[t]
\caption{\label{table:energy}%
The ground-state energy per bond (in unit of $J$) of the spin-1/2
Heisenberg quantum antiferromagnet in the thermodynamic limit for
the linear chain ($d=1$), the square lattice ($d=2$), and the
body-centered-cubic lattice ($d=3$). The results from using FTBC,
FTBC0, TBC or PBC are calculated with $l=2$.}
\begin{ruledtabular}
\begin{tabular}{lllll}
Methods& \;$d=1$ & \;$d=2$ & \;$d=3$ & \;$d=\infty$\\
\hline
exact/best known & $-0.443$\footnotemark[1] & $-0.334$\footnotemark[2] & $-0.288$\footnotemark[3] & $-0.250$ \\
LSW   & $-0.432$ & $-0.329$ & $-0.287$ & $-0.250$\\
FTBC  & $-0.453$ & $-0.332$ & $-0.283$ & $-0.250$\\
FTBC0 & $-0.568$ & $-0.453$ & $-0.379$ & $-0.250$\\
TBC (=PBC) & $-0.750$ & $-0.750$ & $-0.750$ & $-0.750$\\
\end{tabular}
\end{ruledtabular}
\footnotetext[1]{Ref.~\onlinecite{spin:bethe,spin:hulthen}.}%
\footnotetext[2]{Ref.~\onlinecite{spin:square:huse}.}%
\footnotetext[3]{Ref.~\onlinecite{spin:betts_bcc}.}%
\end{table}

\begin{figure}[t]
\begin{center}
\includegraphics[width=0.6\columnwidth,clip=true]{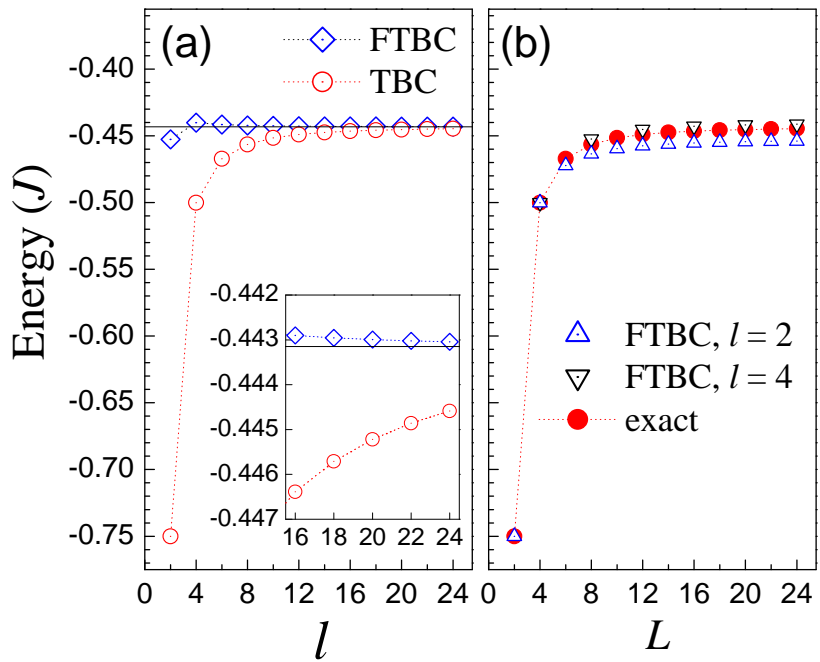}%
\caption{\label{fig:e}%
(a) Bond energy of the one-dimensional spin-1/2 Heisenberg
antiferromagnet for $L \to \infty$ estimated from using FTBC
(diamonds) and TBC or PBC (circles) for a number of $l$. The
horizontal solid line denotes the exact solution ($-J\ln2 + J/4$).
The dotted lines are guide to eyes. (b) Bond energy of the
one-dimensional spin-1/2 Heisenberg antiferromagnet for a number of
$L$ estimated from using FTBC for $l=2$ and $4$ (triangles),
respectively, compared with the exact solutions (circles). }
\end{center}
\end{figure}

Next, the convergency of the bond energy $E(L \to \infty; l)$ with
respect to $l$ is presented in Fig.~\ref{fig:e}(a). The energy of
each subsystem is calculated with the Lanczos exact diagonalization
method and a 5-point Gaussian integration in the twist momentum
space, which is enough for the 6-digit accuracy. It is obvious that
the estimate from FTBC converges to the exact solution $-J\ln 2 +
J/4$ (Refs.~\onlinecite{spin:bethe} and \onlinecite{spin:hulthen})
much faster than the extrapolation to $L \to \infty$ from
finite-size scaling of the PBC or TBC results. Finally, $E(L; 2)$
and $E(L; 4)$ for a number of $L$ are plotted in Fig.
~\ref{fig:e}(b) and compared with the exact solutions for $L$ in
order to show the fast convergence of FTBC with respect to the size
of the simulated system. Overall, the errors for $l=4$ are smaller
than for $l=2$ ($0.003$ versus $0.009$ for $L=24$). For $l=2$, the
small deviations from the exact solutions for $L\ge 4$ follow a
power law $-0.009510(5)+0.152(2)/L^{1.976(6)}$ with
$\chi^2=1.7326\times10^{-11}$. This means that the error grows
rather slowly as $L$ increases away from $l$.   These results
indicate that a large-scale feature of a quantum antiferromagnet
could be captured at the length scale of one lattice constant with
FTBC.


Finally, it is worth mentioning that in the above derivation of
FTBC, we have revealed a less approximated approach, i.e.,
Eq.~(\ref{eq:Q=0}) alone.  It is interesting to explore this
approach and compare it with FTBC. Also, the explicit formulation of
FTBC could facilitate to devise other approximations with lighter
computational load. These studies are beyond the scope of the
present work and will be published elsewhere.

Summarizing, based on coarse graining the space of the momentum
transferred between interacting particles, we have derived the
flavor-twisted boundary condition for simulating quantum many-body
systems with effective Hamiltonians of reduced size. A rapid,
accurate, and fast convergent computation of the ground-state energy
is demonstrated on the spin-$\frac{1}{2}$ quantum antiferromagnet of
any dimension by employing only two sites. The method is expected to
be useful for future simulations and quick estimates on other
strongly correlated systems.


W.Y. is grateful to P. D. Johnson and T. Valla for collaborations
\cite{valla} that stimulated this work. This research utilized
resources at the New York Center for Computational Sciences at Stony
Brook University/Brookhaven National Laboratory which is supported
by the U.S. Department of Energy under Contract No.
DE-AC02-98CH10886 and by the State of New York.



\end{document}